\documentstyle[12pt]{article}
\begin{document}
\title{The Structure of the SWKB Series}
\author{
David T. Barclay \\ 
Dept. of Mathematical Sciences \\
University of Liverpool \\
Liverpool L69 3BX, U.K.}
\date{June 1997} 
\maketitle
\begin{abstract}
The supersymmetric-WKB series is shown to be such that
the SWKB quantisation condition has corrections 
in powers of $\hbar^2$ only and with explicit overall 
factors of $E$.
The results also suggest more efficient methods of
calculating the corrections.
\end{abstract}

\vspace{2.55555in}

\begin{flushright}
LTH 397
\end{flushright}
\newpage

The supersymmetric-WKB approximation is an excellent
example of the way concepts from SUSY quantum mechanics
have illuminated traditional non-relativistic QM
(for a review see \cite{cooper}).
It has provided a new method of estimating energy levels
and wavefunctions which may be superior to
the familiar WKB one \cite{cooper}-\cite{varshni} and it has even
shed new light on the class of exact solutions to the
Schr\"odinger equation \cite{varshni}-\cite{exact}.
Higher-order corrections to the approximation are 
easily calculated \cite{adhikari} and this paper is concerned with 
the structure of these terms.

In SUSY QM one introduces a superpotential $\phi(x)$ and
the operators
\begin{equation}
\label{oper}
A= \hbar {d \over dx} + \phi, \qquad A^+ = - \hbar {d \over dx} + \phi
\end{equation}
so as to define partner Hamiltonians
\begin{equation}
\label{Ham}
H_- = A^+ A, \qquad H_+= AA^+
\end{equation}
which correspond to Schr\"odinger equations (with $2m=1$) involving
the partner potentials
\begin{equation}
\label{pots}
V_- = \phi^2 - \hbar \phi', \qquad V_+ =\phi^2+\hbar\phi
\end{equation}
respectively.
The substitution $\psi= e^{iS/\hbar}$ allows the Schr\"odinger
equation for $V_-$ to be written as
\begin{equation}
\label{ricc}
{S'}^2 - i \hbar S'' + \phi^2-\hbar \phi' =E.
\end{equation}
A formal series solution 
\begin{equation}
\label{series}
S'(x) = \sum_{n=0}^{\infty} (-i\hbar)^n {S_n}'(x)
\end{equation}
starting with  
\begin{equation}
\label{s1s2}
{S_0}' = (E-\phi^2)^{1/2}, \qquad
{S_1}' = {\phi\phi' \over 2(E-\phi^2)} +
   {i \phi' \over 2 (E-\phi^2)^{1/2}}
\end{equation}
can be generated recursively and is called the SWKB series.
Its main use is in the SWKB quantisation condition
\begin{equation}
\label{quan}
\oint \sum_{n=0}^{\infty} (-i\hbar)^n {S_n}'(x) dx
 = 2 (n+ 1/2) \pi \hbar.
\end{equation}
This has previously been calculated to $o(\hbar^6)$
\cite{adhikari}, 
the first few terms being
\begin{eqnarray}
\oint (E&-&\phi^2)^{1/2} dx  -  {3\hbar^2 E \over 24} \oint
{{\phi'}^2 \over (E-\phi^2)^{5/2}}dx \nonumber \\
& - & {\hbar^4 E \over128} \oint \biggl( {49 E {\phi'}^4
\over (E-\phi^2)^{11/2}} - {140 \over 3} {{\phi'}^4
\over (E-\phi^2)^{9/2}} - {4\phi' {\phi'''} \over
(E-\phi^2)^{7/2}} \biggr) dx \nonumber \\ 
& + & \ldots = 2n \pi \hbar
\label{cond}
\end{eqnarray}
following simplification using integration by parts.

Note that although the ${S_n}'$ are in general part imaginary,
(\ref{cond}) will be entirely real once the contour integral is
collapsed down into one along the real axis.
Apart from being necessary physically, this property of the 
series is readily proved directly.
If $S' = R +iI$, then dividing (\ref{ricc}) into real and
imaginary parts gives
\begin{equation}
\label{imag}
I = {\hbar \over 2}{d \over dx} ( \ln R),
\end{equation}
which is zero after integration (c.f. \cite{capri}\cite{vasan}).
It will also become convenient to split ${S_n}'$ into real
and imaginary parts thus
\begin{equation}
\label{split}
{S_n}' = p_n + i q_n.
\end{equation}
In this notation, (\ref{imag}) demonstrates that
\begin{equation}
\label{comzero}
\oint p_{2n+1} dx = \oint q_{2n} dx =0.
\end{equation}

Several other patterns ought to be noted in (\ref{cond}).
Apart from the $o(\hbar)$ term ${S_1}'$, which is finite and
exactly cancels the extra $\pi \hbar/2$ in (\ref{quan}), all
terms involving odd powers of $\hbar$ have vanished in the
quantisation condition integral.
For example
\begin{equation}
\label{q3}
\oint q_3 dx = {1 \over 16} \oint {d \over dx}\biggl[
{5 \phi {\phi'}^2 \over (E-\phi^2)^{5/2}}
+ {2 \phi'' \over (E-\phi^2)^{3/2}} \biggr] dx = 0.
\end{equation}
One can conjecture \cite{adhikari}\cite{vasan} that
all the $q_{2n+1}$ terms can similarly be written as
derivatives and hence that
\begin{equation}
\label{conj}
\oint q_{2n+1} dx = 0, \qquad \forall n>0.
\end{equation}
This has been confirmed by direct calculation 
up to $n=2$ in general \cite{adhikari}
 and up to $n=5$ for the
particular case of $\phi = x^{2N+1}/(2N+1)$
\cite{vasan}.
The equivalent WKB result is known to be true to all
orders in $\hbar$, but this follows from the fact
that that $i$ and $\hbar$ are paired in the WKB 
version of (\ref{ricc}) and so the requirement 
that the quantisation condition be real already 
covers this \cite{capri}.
Surprisingly, while formally so similar, the
SWKB result will derive from an entirely different
symmetry. 
 
Furthermore, even the terms that do not vanish in the 
integration, i.e. the $p_{2n}$, appear to be such that 
the eventual corrections are proportional to $E$.
While it is true that this property elegantly accords 
with the fact that the lowest-order SWKB condition is exact
for ground states (for which $E=0$ in SUSY QM)
\cite{comtet}, it does
not follow from it: as happens in the lowest-order case,
the integrals themselves could tend to
zero when $E \rightarrow 0$
without an explicit overall factor of $E$.

Before proving the first of these properties, we note a 
consequence of it that is closely related to the proof.
One well-known theorem in SUSY QM is that the eigenvalues
of $V_-$ and $V_+$ satisfy ${E_n}^{(-)} = {E_{n-1}}^{(+)}$.
To $o(\hbar)$ the SWKB approximation applied to $V_+$ gives a
result equivalent to the $V_-$ one (\ref{cond}), except that
the $n$ on the right hand side is replaced by
$n-1$.
The lowest-order SWKB estimates for the eigenvalues thus
preserve the degeneracy between the spectra \cite{dutt}.
It is natural to conjecture that this remains true when the
approximation is truncated at higher-orders, but for this to
be true requires (\ref{conj}).
Specifically, if the series solution to
\begin{equation}
\label{ricc+}
{{S^{(+)}}'}^2 - i \hbar {S^{(+)}}'' + \phi^2 + \hbar \phi'=E,
\end{equation}
the $V_+$ equivalent of (\ref{ricc}), is compared to the $V_-$
version, one finds using the substitution $\hbar \rightarrow -\hbar$,
$i \rightarrow -i$ that 
\begin{equation}
\label{diff}
{{S_n}^{(+)}}' = {S_n}' - 2i q_n.
\end{equation}
Any non-zero $q_n$ integrals spoil the symmetry between the $V_-$
and $V_+$ quantisation conditions.
Again this only makes a conjecture extremely natural, without proving it.

To prove it, consider the $\psi^{(-)} = e^{iS/\hbar}$ and 
$\psi^{(+)}=e^{iS^{(+)}/\hbar}$ which led to (\ref{ricc}) 
and (\ref{ricc+})
respectively.
These are solutions to
\begin{equation}
\label{eigen}
H_- \psi^{(-)} = E \psi^{(-)}, \qquad H_+ \psi^{(+)}
= E \psi^{(+)}.
\end{equation}
As always in SUSY QM, the operators (\ref{oper}) relate eigenfunctions
of partner Hamiltonians and in particular
\begin{equation}
\label{raise}
 \psi^{(+)} = A \psi^{(-)},
\end{equation}
which directly implies that
\begin{equation}
\label{answer}
{S^{(+)}}' = S' - i\hbar {d \over dx} \ln (\phi +iS')
\end{equation}
relates the two solutions.

Strictly however the operator algebra only shows that $A\psi^{(-)}$
is, like $\psi^{(+)}$, a solution to $H_+ \psi = E \psi$.
But there are infinitely many solutions.
In principle this derivation of (\ref{answer}) implicitly assumes
that the series solution (\ref{series}) already defines a 
boundary condition for the wavefunction $\psi^{(-)}
= e^{iS/\hbar}$ and also for $\psi^{(+)}$ such that (\ref{raise})
is true.
Since the series (\ref{series}) is presumably divergent, this is
at best delicate.
However given (\ref{answer}) one can directly show that it is a 
solution to (\ref{ricc+}).
It must now be true  {\it order-by-order} in $\hbar$ that the 
solution (\ref{series}) to (\ref{ricc}) implies a solution 
(\ref{answer}) that is the unique solution to (\ref{ricc+}) to
that order in $\hbar$.
Put another way, if $A \psi^{(-)}$ and $\psi^{(+)}$ are 
different solutions, they are still related in such a way that
(\ref{answer}) can only omit contributions that are 
non-perturbative in $\hbar$.
It is thus adequate for comparing SWKB series order-by-order.

Given the SWKB series for $V_-$ one can now find that for $V_+$ 
by expanding the right hand side of (\ref{answer}) as a power
series in $\hbar$.
And because of (\ref{diff}) one finds that
\begin{equation}
\label{total}
q_{n+1} = {i \over 2}{d \over dx}(L_n)
\end{equation}
where
\begin{equation}
\label{L0}
L_0 = \ln (\phi + i \sqrt{ E - \phi^2})
\end{equation}
\begin{equation}
\label{Ln}
L_n = i (\phi+i \sqrt{E-\phi^2})^{-1}
\biggl[ {S_n}' - \sum_{m=0}^{n-2} (m+1) L_{m+1}
	{S'}_{n-1-m} \biggr].
\end{equation}
The logarithm in $L_0$ means that $q_1$ integrates to $\pi \hbar$
as required, but all the higher $q_n$ are total derivatives which
can be eliminated using integration by parts, 
thereby proving (\ref{conj}).

This result can now be used to prove the conjecture relating to
the $p_{2n}$.
To do so, note that the  
$p_n$ and $q_n$ are more closely related
than they first appear.
One significant difference between them is that whereas in any order
one member of the pair contains denominators made up of odd powers of
$(E-\phi^2)^{1/2}$, in the same order the other contains only even 
powers.
Compensating for this by introducing a factor $F \equiv \phi/
(E-\phi^2)^{1/2}$, one discovers that
\begin{equation}
\label{decom}
p_n=Fq_n+E \alpha_n, \qquad q_n=-Fp_n + E \beta_n,
\end{equation}
starting with
\begin{equation}
\label{first}
p_1=Fq_1, \qquad q_1 =-Fp_1 + {E\phi' \over 2
(E-\phi^2)^{3/2}}.
\end{equation}
The general result is easily proved by induction using the
recurrence relations for $p_n$ and $q_n$ that can be
derived from (\ref{ricc}).
One of the reasons this proof works is that 
$F'=E\phi'/(E-\phi^2)^{3/2}$.
For the same reason
\begin{eqnarray}
\oint p_{2n} dx & = & \oint \biggl[ {d \over dx} (FQ_{2n})
-F' Q_{2n} + E \alpha_{2n} \biggr] dx \nonumber \\
& = & E \oint \biggl[ \alpha_{2n} - {\phi' Q_{2n} \over
(E-\phi^2)^{3/2}}\biggr] dx, \label{fac}
\end{eqnarray}
where $Q_{2n}= i L_{2n-1}/2$, 
such that ${Q_{2n}}'= q_{2n}$, 
is now known to exist.
The origin of the overall $E$ factor in the corrections
in (\ref{cond}) thus becomes clear.

These new results (\ref{Ln}) and (\ref{fac}) can be exploited
to make any  future attempts to calculate further corrections
to (\ref{cond}) more efficient.
While the ${S_n}'$ are easily found recursively, the multiple
integrations by parts necessary to maximally simplify the
quantisation condition are difficult to specify algorithmically.
But (\ref{fac}) now indicates that these are equivalent to the
much simpler operation of subtracting $(FQ_{2n})'$ from $p_{2n}$;
this leaves a simpler correction with an overall factor of $E$.

As a varient of this, introduce $P(x)\equiv \sum_{n=0}^{\infty}
(-i\hbar)^n p_n$ and its $q_n$ equivalent. 
Because of (\ref{ricc}), these functions obey
\begin{equation}
\label{sys}
-i\hbar P' = -P^2 +Q^2 + {p_0}^2, \qquad
i \hbar Q' = 2PQ + i \hbar \phi',
\end{equation}
where $p_0=(E-\phi^2)^{1/2}$, and also, because of (\ref{decom})
\begin{equation}
\label{link}
P = p_0 + FQ + E \alpha (x).
\end{equation}
In the $E\rightarrow 0$ limit, this system reduces to
$P=i \phi - iQ$ and
\begin{equation}
\label{Ezero}
 -i \hbar P' = -2P^2 + 2iP \phi,
\end{equation}
with solution
\begin{equation}
\label{zerosol}
P = i \phi + {i \hbar \over 2} {d \over dx}( \ln P).
\end{equation}
Now consider a function $\overline P$ defined via
\begin{equation}
\label{pbar}
\overline P = \sqrt{E-\phi^2} + {i \hbar \over 2}
{d \over dx}(\ln \overline P ).
\end{equation}
This clearly has the same $E \rightarrow 0$ limit as $P$ and
indeed $\overline P$ ought to be thought of as containing 
the terms in $P$ that do not vanish in this limit.
Furthermore its ${\overline p}_n$ will all be 
writable as total derivatives for $n>0$.
Thus a strategy for simplifying the corrections in (\ref{cond})
is to calculate the $p_n$ and $q_n$ as normal, but in parallel
calculate the ${\overline p}_n$ implied by (\ref{pbar}).
Using $(p_{2n}-{\overline p}_{2n})$ in the quantisation condition
is then equivalent to using $p_{2n}$, but 
again the subtraction eliminates
the terms normally removed using integration by parts.

Finally, as an aside, we note that if the standard WKB quantisation
condition has already been calculated to some order, the SWKB
one can be found to the same order without having to calculate
the full SWKB series. 
As is well-known \cite{adhikari}, an alternative 
to using recurrence relations
to calculate (\ref{cond}) is to make the substitution 
$V=\phi^2-\hbar \phi'$ in the WKB series and re-expand in $\hbar$.
However if it is only the quantisation condition that is of
interest, the substitution can be made at this level,
i.e. after the WKB condition has been simplified.
This is possible because the terms eliminated in going from
the WKB series to the WKB quantisation condition can still be
written as derivatives after the substitution and re-expansion
and their contribution to the SWKB quantisation condition would
thus be zero anyway.
Also
\begin{eqnarray}
\oint {V' \over E-V}dx & = & \oint {2\phi \phi' \over E-\phi^2}dx
+ \sum_{n=1}^{\infty}{(-\hbar)^n \over n} \oint
{d \over dx} \biggl( {\phi' \over E-\phi^2} \biggr)^n dx
\nonumber \\
& = & \oint {2 \phi \phi' \over E - \phi^2 } dx.
\label{ohbar}
\end{eqnarray}
The $o(\hbar)$ term in the WKB series that produces the constant
in the quantisation condition thus just gives an $o(\hbar)$ 
constant in the SWKB quantisation condition, as required.
The main disadvantage of this method of deriving (\ref{cond})
-- apart from the inconvenient fact that the WKB quantisation
condition is not currently known to higher than 
$\hbar^6$ \cite{krieger}\cite{adhikari} --
is that the result will not be fully simplified, yet neither
can the simplification methods proposed above be used here.

That the SWKB series is structured in the ways implied by
(\ref{Ln}), (\ref{decom}) and (\ref{fac}) proves all extant conjectures
about the form of the SWKB quantisation condition.
(\ref{Ln}) in particular is a direct consequence of the 
supersymmetry relating $H_-$ and $H_+$.
Whether any further patterns exist in the series remains to
be discovered.

$$ $$

{\bf Acknowledgements}
$$ $$

Thanks to Adrian Campbell-Smith for helpful empirical 
observations and to Chris Maxwell for passing them on.


\newpage
\begin{thebibliography}{19}
\renewcommand{\baselinestretch}{1.0}

\bibitem{cooper} F. Cooper, A. Khare and U.P. Sukhatme,
		 Phys.Rep.{\bf 251} (1995) 267.

\bibitem{khare} A. Khare, Phys.Lett.{\bf B161} (1985) 131;
		R. Dutt, A. Khare and Y.P. Varshni,
		Phys.Lett.{\bf A123} (1987) 375;
		P. Roy, R. Roychoudhuri and Y.P. Varshni,
		J.Phys.{\bf A21} (1988) 1587;
		D. Delaney and M.M. Nieto,
		Phys.Lett.{\bf B247} (1990) 301;
		Y.P. Varshni, J.Phys.{\bf A25} (1992) 5761;
		M. Hr\v uska, W.-Y. Keung and U.P. Sukhatme,
		UIC preprint UICHEP-TH/96-18, quant-ph/9611030.

\bibitem{varshni} A. Khare and Y.P. Varshni, Phys.Lett.{\bf A142} 1. 

\bibitem{dutt} R. Dutt, A. Khare and U.P. Sukhatme, 
	        Phys.Lett.{\bf B181} (1986) 295.

\bibitem{exact} K. Raghunathan, M. Seetharaman and S.S. Vasan,
		Phys.Lett.{\bf B188} (1987) 351;
		D. Barclay and C.J. Maxwell,
		Phys.Lett.{\bf A157} (1991) 351.

\bibitem{adhikari} R. Adhikari, R. Dutt, A. Khare and U.P. Sukhatme,
                   Phys.Rev.{\bf A38} (1988) 1679.  

\bibitem{capri} J. Dunham, Phys.Rev.{\bf 41} (1932) 721;
		C.M. Bender, K. Olaussen and P.S. Wang,
		Phys.Rev.{\bf D16} (1977) 1740;
		A.Z. Capri, {\it Nonrelativistic Quantum Mechanics},
		Benjamin/Cummings, 1985, p391.

\bibitem{vasan} S.S. Vasan, M. Seetharaman and K. Raghunathan,
		J.Phys.{\bf A21} (1988) 1897.

\bibitem{comtet} A. Comtet, A.D. Bandrauk and D.K. Campbell,
		 Phys.Lett.{\bf 150B} (1985) 159.

\bibitem{krieger} J.B. Krieger, M.L. Lewis and C. Rozenzweig,
		  J.Chem.Phys.{\bf 47} (1967) 2942.



\end{thebibliography}
\end{document}